\documentclass[crystals,article,submit,moreauthors,pdftex,12pt,a4paper]{mdpi} 
\lastpage{x}
\doinum{10.3390/------}
\pubvolume{xx}
\pubyear{2012}
\history{Received:  / Accepted:  / Published: }
\usepackage{graphicx}
\Title{ Bound States and Supercriticality
in Graphene-Based Topological Insulators }

\Author{Denis Kl\"opfer $^{1}$, Alessandro De Martino $^{2}$ 
and Reinhold Egger $^{1,\star}$}

\address{
$^{1}$ Institut f\"ur Theoretische Physik, Heinrich-Heine-Universit\"at,
 D-40225 D\"usseldorf, Germany\\
$^{2}$ Department of Mathematical Science,
City University London, London EC1V 0HB, United Kingdom}

\corres{Email: egger@thphy.uni-duesseldorf.de, Tel. (49) 211 81-14710, 
Fax (49) 211 81-15630.}

\abstract{We study the bound state spectrum and the conditions for 
entering a supercritical regime in 
graphene with strong intrinsic  and Rashba 
spin-orbit interactions within the topological insulator phase.
Explicit results are provided for a disk-shaped potential well and 
for the Coulomb center problem. 
 }


\begin{document}

\section{Introduction}

The electronic properties of graphene monolayers are presently under intense
study. Previous works have already revealed many novel and fundamental insights;
for reviews, see \cite{castro,kotov}.  Following the seminal work of 
Kane and Mele \cite{kane}, it may be possible to engineer a 
two-dimensional (2D) topological insulator (TI) phase \cite{hasan} 
in graphene by enhancing the --- usually very
weak \cite{brataas,alan,fang} --- spin-orbit interaction (SOI) in graphene. 
This enhancement could, for instance, be achieved by the deposition of 
suitable adatoms \cite{weeks}.  Remarkably, random deposition should 
already be sufficient to reach the TI phase \cite{waintal1,waintal2,jiang}
where the effective ``intrinsic'' SOI $\Delta$ 
exceeds (half of) the ``Rashba'' SOI $\lambda$.  
So far, the only 2D TIs realized experimentally are based
on the mercury telluride class.  Using graphene
as a TI material constitutes a very attractive option because of 
the ready availability of high-quality graphene samples \cite{castro}
and the exciting prospects for  stable and robust 
TI-based devices \cite{hasan}, see also \cite{dario,lenz}   

In this work, we study bound-state solutions and the conditions for 
supercriticality in a graphene-based TI.  
Such questions can arise in the presence of an electrostatically
generated potential well (``quantum dot'') or for a Coulomb center.
The latter case can be realized by artificial alignment of 
Co trimers \cite{exp}, or when defects or charged impurities reside
in the graphene layer.
Without SOI, 
the Coulomb impurity problem in graphene has been theoretically 
studied in depth 
\cite{katsnelson,pereira,shytov,shytov2,biswas,fogler}; for 
reviews, see \cite{castro,kotov}.
Moreover, for $\lambda=0$, an additional mass term in the Hamiltonian
corresponds to the intrinsic SOI $\Delta$ (see below), and the massive
Coulomb impurity problem in graphene has been analyzed in 
\cite{novikov,terekhov,pereira2,gamayun,gamayun2,zhu}.
However, a finite Rashba SOI $\lambda$ is inevitable in practice 
and has profound consequences. In particular, 
$\lambda\ne 0$ breaks electron-hole symmetry and
modifies the structure of the vacuum.  
We therefore address the general case with both $\Delta$ and
$\lambda$ finite, but within the TI phase $\Delta>\lambda/2$, in this
paper.
Experimental progress on the observation of Dirac quasiparticles near
a Coulomb impurity in graphene was also reported very recently \cite{exp}, and
we are confident that the topological version with enhanced SOI 
can be studied experimentally in the near future.
Our work may also be helpful in the understanding
of spin-orbit mediated spin relaxation in graphene \cite{brataas2}. 

The atomic collapse problem for Dirac fermions in an attractive Coulomb 
potential,
$V(r)=-\hbar v_F \alpha/r$,   
could thereby be realized in topological graphene. 
Here we use the dimensionless impurity strength 
\begin{equation}\label{alphadef}
\alpha=\frac{Ze^2}{\kappa \hbar v_F}\simeq 2.2 \frac{Z}{\kappa} ,
\end{equation}
where $Z$ is the number of positive 
charges held by the impurity, $\kappa$ a dielectric constant 
characterizing the environment, and 
$v_F\approx 10^6$~m$/$s the Fermi velocity.
Without SOI, the Hamiltonian is not self-adjoint for $\alpha>\alpha_c=1/2$, 
and the 
potential needs short-distance regularization,
e.g., by setting $V(r<R)=-\hbar v_F\alpha/R$ with 
short-distance cutoff $R$  of the order of the lattice constant of graphene
\cite{castro,kotov}.
Including a finite ``mass'' $\Delta$, i.e., the intrinsic SOI,
but keeping $\lambda=0$, the critical coupling $\alpha_c$ is shifted to
\cite{gamayun}
\begin{equation}\label{alphac} 
\alpha_c\simeq \frac12+\frac{ \pi^2}{\ln^2(0.21\Delta R/\hbar v_F)},
\end{equation}
approaching the value $\alpha_c=1/2$ for $R\to 0$.
In the supercritical regime $\alpha>\alpha_c$, the lowest bound 
state ``dives'' into
the valence band continuum (Dirac sea).  It then
becomes a resonance with complex energy, where the imaginary
part corresponds to the finite decay rate into the continuum.
Below we show that the Rashba SOI provides an interesting
twist to this supercriticality story.
 
The structure of this article is as follows.
In Sec.~\ref{sec2} we introduce the model and summarize its symmetries.
The case of a circular potential well is addressed in Sec.~\ref{sec3}
before turning to the Coulomb center in Sec.~\ref{sec4}.
Some conclusions are offered in Sec.~\ref{sec5}.
Note that we do not include a magnetic field (see, e.g., \cite{rashba,ademarti})
and thus our model enjoys time-reversal symmetry.
Below, we often use units with $\hbar=v_F=1$.

\section{Model and symmetries}\label{sec2}

\subsection{Kane-Mele model with radially symmetric potential}

We study the Kane-Mele model for a 2D graphene monolayer with 
both intrinsic ($\Delta$) and Rashba ($\lambda$)  SOI \cite{kane}
in the presence of a radially symmetric scalar potential $V(r)$. 
Assuming that $V(r)$ is sufficiently smooth to allow for the neglect of 
inter-valley scattering, the 
low-energy Hamiltonian near the $K$ point $(\tau=+1)$ is given by
\begin{equation}\label{km-model}
H =  \tau \sigma_x p_x + \sigma_y p_y + \tau \Delta \sigma_z s_z
+\frac{\lambda}{2} (\tau \sigma_x s_y-\sigma_y s_x) + V(r) ,
\end{equation}
with Pauli matrices $\sigma_{x,y,z}$ ($s_{x,y,z}$) in sublattice (spin) 
space \cite{castro}.
The Hamiltonian near the other valley ($K'$ point) follows for
$\tau=-1$ in Eq.~(\ref{km-model}).
We note that a sign change of the Rashba SOI, $\lambda\to -\lambda$,
does not affect the spectrum due to the relation $H(-\lambda)=
s_z H(\lambda) s_z$.  Without loss of generality, we then
put $\Delta\ge 0$ and $\lambda\ge 0$.

Using polar coordinates, it is now straightforward to verify 
(see also \cite{novikov}) that total angular momentum, defined as
\begin{equation}
J_z = -i\partial_\phi+ s_z/2+ \tau\sigma_z/2, 
\end{equation}
is conserved and has integer eigenvalues $j$. For given $j$,
eigenfunctions of $H$ must then be of the form 
\begin{equation}
\Psi_{j,\tau=+}(r,\phi) =  \left( \begin{array}{c} 
e^{i(j-1)\phi}\ a_{\uparrow,j,+}(r)\\ ie^{ij\phi}\ b_{\uparrow,j,+}(r)\\
e^{ij\phi} \ a_{\downarrow,j,+}(r)\\ i e^{i(j+1)\phi} \
b_{\downarrow,j,+}(r)\end{array}\right),
\quad \Psi_{j,-}(r,\phi) =  \left( \begin{array}{c} 
e^{ij\phi}\ a_{\uparrow,j,-}(r)\\ ie^{i(j-1)\phi}\ b_{\uparrow,j,-}(r)\\
e^{i(j+1)\phi} \ a_{\downarrow,j,-}(r)\\ i e^{ij\phi} \
b_{\downarrow,j,-}(r)\end{array}\right).
\end{equation}
Next we combine the radial functions to (normalized) four-spinors,
\begin{equation}\label{4spinor}
\Phi_{j,\tau}(r)=\left( \begin{array}{c} a_{\uparrow,j,\tau}(r) 
\\ b_{\uparrow,j,\tau}(r)\\ 
a_{\downarrow,j,\tau}(r) \\ b_{\downarrow,j,\tau}(r) \end{array}\right).
\end{equation}
In this representation, the radial Dirac equation 
for total angular momentum $j$ and valley index $\tau=\pm$ reads
\begin{equation}\label{dirac} 
( H_{j,\tau} - E ) \ \Phi_{j,\tau}(r) =0,
\end{equation}
with Hermitian matrix operators (note that $\Delta$ denotes
the intrinsic SOI and not the Laplacian)
\begin{eqnarray}\label{dirac2}
 H_{j,+} &=& \left( \begin{array}{cccc} 
\Delta+V & \nabla^{(+)}_j & 0 &0 \\
\nabla_{j-1}^{(-)}  & -\Delta+V &-\lambda  & 0 \\
0 & -\lambda &  -\Delta+V & \nabla^{(+)}_{j+1} \\
0 &  0 & \nabla^{(-)}_j & \Delta+V \end{array} \right) ,
 \\ \nonumber H_{j,-} &=& \left( \begin{array}{cccc} 
-\Delta+V & \nabla^{(-)}_{j-1} & 0 &-\lambda \\
\nabla_{j}^{(+)}  & \Delta+V & 0 & 0 \\
0 & 0 &  \Delta+V & \nabla^{(-)}_{j} \\
-\lambda &  0 & \nabla^{(+)}_{j+1} &- \Delta+V \end{array} \right) ,
\end{eqnarray}
where we use the notation 
\begin{equation}\label{nabla}
\nabla^{(\pm)}_j=\frac{j}{r} \pm \frac{d}{dr}.
\end{equation}

One easily checks that Eq.~(\ref{dirac2}) satisfies the 
parity symmetry relation
\begin{equation}\label{sym1}
H_{-j,\tau}= \sigma_y s_y H_{j,\tau} \sigma_y s_y.
\end{equation}
Note that this ``parity'' operation for the radial Hamiltonian
is non-standard in the sense that the valley is not changed by
the transformation $\sigma_y s_y$, 
spin and sublattice are flipped simultaneously, and
only the $y$-coordinate is reversed.  (We will nonetheless refer to
$\sigma_y s_y$ as parity transformation below.) 
A second symmetry relation connects both valleys,
\begin{equation}\label{sym2}
H_{j,-\tau}= \sigma_x H_{j,\tau} \sigma_x.
\end{equation}
Using Eq.~(\ref{sym1}), this relation can be traced back to a 
time-reversal operation.
Equations (\ref{sym1}) and (\ref{sym2}) suggest that eigenenergies 
typically are four-fold degenerate.

When projected to the subspace of fixed (integer) total angular momentum $j$, 
the current density operator has angular component $J_{\phi}= \sigma_x$ 
and radial component $J_r=-\tau \sigma_y$ for arbitrary $j$. 
When real-valued entries can be chosen in $\Phi_{j,\tau}(r)$, 
the radial current density thus vanishes separately in each valley.
We define the (angular) spin current density as 
$J^S_\phi=s_z\sigma_x$.  Remarkably, the transformation defined in
Eq.~(\ref{sym2}) conserves both (total and spin) angular currents,
while the transformation in 
Eq.~(\ref{sym1}) reverses the total current but conserves the 
spin current.  Therefore, at any energy, 
eigenstates supporting spin-filtered counterpropagating currents are 
possible.
However, in contrast to the edge states found in a ribbon geometry
\cite{kane}, these spin-filtered states do not necessarily have a 
topological origin. 

We focus on one $K$ point ($\tau=+$) and omit the $\tau$-index henceforth; 
the degenerate $\tau=-$ Kramers partner easily follows using Eq.~(\ref{sym2}).  
In addition, using the symmetry (\ref{sym1}), it is sufficient to study
the model for fixed total angular momentum $j\ge 0$.

\subsection{Zero total angular momentum}
\label{sec22}

For arbitrary $V(r)$, we now show that a drastic simplification
is possible for total angular momentum $j=0$, which
can even allow for an exact solution.  Although
the lowest-lying bound states for the potentials in Secs.~\ref{sec3}
and \ref{sec4} are found in the $j=1$ sector, exact statements
about what happens for $j=0$ are valuable and can be explored along
the route sketched here.

The reason why $j=0$ is special can be seen from the parity symmetry
relation in Eq.~(\ref{sym1}). The parity transformation $\sigma_y s_y$ 
connects the $\pm j$ sectors, but represents a discrete symmetry of 
the $j=0$ radial Hamiltonian $H_{j=0,\tau}$
[see Eq.~(\ref{dirac2})]
acting on the four-spinors in Eq.~(\ref{4spinor}). 
Therefore, the $j=0$ subspace can be 
decomposed into two orthogonal subspaces corresponding to the two distinct
eigenvalues of the Hermitian operator $\sigma_y s_y$.
This operator is diagonalized by the matrix
\begin{equation}\label{unitary}
U = \frac{1}{\sqrt{2}} \left(\begin{array}{cccc} 1& 0 & 0 & -1\\
0 & 1 & 1 & 0\\  1 & 0 & 0 & 1 \\  0 & 1 &  -1&0 \end{array}\right),
\end{equation}
such that 
\begin{equation}\label{diagonl}
U \sigma_y s_y U^{-1} = {\rm diag}(1,1,-1,-1).
\end{equation}
In fact, using this transformation matrix to carry out a similarity 
transformation, $\tilde H_{j,+}=UH_{j,+} U^{-1}$, we obtain 
\begin{equation}\label{tildeh0}
\tilde H_{j,+} = \left( \begin{array}{cccc} 
\Delta+V & \partial_r & 0 & j/r \\
-1/r-\partial_r  & -\Delta+V-\lambda & j/r & 0 \\
0& j/r  &  \Delta+V & \partial_r  \\
j/r & 0 & -1/r -\partial_r& -\Delta+\lambda+V \end{array} \right). 
\end{equation}
For $j=0$, the upper and lower $2\times 2$ blocks decouple.
Each block has the signature (``parity'') $\sigma=\pm$
corresponding to the eigenvalues in Eq.~(\ref{diagonl}), and
represents a mixed sublattice-spin state, see Eqs.~(\ref{4spinor}) and
 (\ref{unitary}). 

For parity $\sigma=\pm$, the $2\times 2$ block
matrix in Eq.~(\ref{tildeh0}) is formally identical to an effective $\lambda=0$ 
problem with $j=0$, fixed $s_z=\sigma$, and the substitutions 
\begin{equation}\label{subst}
\Delta\to \Delta+ \sigma\lambda/2,\quad E\to E+\sigma \lambda/2.
\end{equation}
This implies that for $j=0$ and arbitrary $V(r)$, 
the complete spectral information 
for the full Kane-Mele problem (with $\lambda\ne 0$) directly follows
from the $\lambda=0$ solution.

\subsection{Solution in region with constant potential}

We start our analysis of the Hamiltonian (\ref{km-model}) 
with the general solution of Eq.~(\ref{dirac}) 
for a region of constant potential.  Here, it 
suffices to study $V(r)=0$, since $E$ and $V$ enter only
through the combination $E-V$ in Eq.~(\ref{dirac2}). 
In Sec.~\ref{sec3}, we will use this solution
to solve the case of a step potential.  

The general solution to Eq.~(\ref{dirac}) follows from the Ansatz
\begin{equation}\label{generalfree1}
\Phi_j(r)= \left(\begin{array}{l} c_1 B_{j-1}(\sqrt{p}r)\\
c_2 B_{j}(\sqrt{p}r) \\
c_3 B_{j}(\sqrt{p}r) \\
c_4 B_{j+1}(\sqrt{p}r) \end{array}\right).
\end{equation}
where the $c_i$ are real coefficients,
$B_j$ is one of the cylinder (Bessel) functions, $B_j=J_j$ or $B_j=H^{(1)}_j$,
and $p$ denotes a real spectral parameter. 
In particular, $\sqrt{p}$ is a generalized radial wavenumber.
We here assume true bound-state solutions with real-valued energy.
However, for quasi-stationary resonance states with complex energy, 
$p$ and the $c_i$ may be complex as well.

Using the Bessel function recurrence relation,
$\nabla^{(\pm)}_j B_{j}(\sqrt{p}r)=\sqrt{p} B_{j\mp 1}(\sqrt{p}r)$, 
the set of four coupled differential equations 
(\ref{dirac}) simplifies to a set of algebraic equations,
\begin{equation}\label{alg}
\left( \begin{array}{cccc} \Delta- E & \sqrt{p} & 0 &0 \\
\sqrt{p} & -\Delta-E &-\lambda  & 0 \\
0 & -\lambda &  -\Delta-E & \sqrt{p} \\
0 &  0 & \sqrt{p} & \Delta-E \end{array} \right) 
\left(\begin{array}{c} c_1\\ c_2\\ c_3 \\ c_4\end{array}\right) = 0.
\end{equation}
Notably, $j$ does not appear here, and therefore
the spectral parameter $p$ depends only on the energy $E$.
The condition of vanishing determinant then yields a quadratic equation for
$p$, with the two solutions 
\begin{equation}\label{ppm}
p_\pm = (E-\Delta)(E-E_\pm), \quad
E_\pm=-\Delta\pm \lambda.
\end{equation}
Which Bessel function is chosen in Eq.~(\ref{generalfree1}) 
now depends on the sign of $p_\pm$
and on the imposed regularity conditions for $r\to 0$ and/or $r\to \infty$.

For $p_\pm > 0$, a solution  regular at the origin is 
obtained by putting $B_j=J_j$, 
which describes standing radial waves. 
Equation (\ref{alg}) then yields the unnormalized spinor 
\begin{equation}\label{generalfree2}
\Phi_{j,p_\pm>0} (r)= \left(\begin{array}{c} 
\frac{\sqrt{p_\pm}}{E-\Delta} J_{j-1}(\sqrt{p_\pm}r) \\ 
J_{j}(\sqrt{p_\pm}r) \\  
\mp J_{j}(\sqrt{p_\pm}r) \\
\mp \frac{\sqrt{p_\pm}}{E-\Delta} J_{j+1}(\sqrt{p_\pm} r) 
\end{array}\right).
\end{equation}

For $p_\pm<0$, instead it is convenient to set $B_j =H^{(1)}_j$ 
in Eq.~(\ref{generalfree1}).
Using the identity $H^{(1)}_j(ze^{i\pi/2})=\frac{2}{\pi i}e^{-ij\pi/2}K_j(z)$, 
the unnormalized spinor resulting from Eq.~(\ref{alg})
then takes the form
\begin{equation} \label{Keigenstate}
\Phi_{j,p_\pm<0} (r)= \left(\begin{array}{c} 
-\frac{\sqrt{-p_\pm}}{E-\Delta} K_{j-1}(\sqrt{-p_\pm} r) \\ 
 K_{j}(\sqrt{-p_\pm}r) \\  
\mp K_{j}(\sqrt{-p_\pm}r) \\
\mp \frac{\sqrt{-p_\pm}}{E-\Delta} K_{j+1}(\sqrt{-p_\pm} r) 
\end{array}\right), 
\end{equation}
where the modified Bessel function $K_j(\sqrt{-p_\pm} r)$ 
describes evanescent modes, exponentially decaying 
at infinity.

\subsection{Solution without potential}

In a free system, i.e., when $V(r)=0$ for all $r$, the only acceptable 
solution corresponding to a physical state is obtained when $p_\pm>0$ 
\cite{rakyta}. 
For $\Delta<\lambda/2$, at least one $p_\pm>0$ in 
Eq.~(\ref{ppm}) for all $E$, and the system is gapless.
However, the TI phase defined by $\Delta>\lambda/2$ 
has a gap as we show now.

For $\Delta>\lambda/2$, Eq.~(\ref{ppm}) tells us
that for $E>\Delta$ and for $E<E_-$, both solutions 
$p_\pm$ are positive and hence (for given $j$ and $\tau$)
there are two eigenstates $\Phi_{j,p_\pm}$ for given energy $E$.
However, within the energy window [with $E_\pm$ in Eq.~(\ref{ppm})]
\begin{equation}\label{window2}
E_-<E<E_+,
\end{equation}
we have $p_+>0$ and $p_-<0$, i.e., only the eigenstate $\Phi_{j,p_+}$
represents a physical solution. Both $p_\pm$ are negative when
$E_+<E<\Delta$, and no physical state exists at all.
This precisely corresponds to the topological gap in 
the TI phase \cite{kane}.  Note that due to the Rashba SOI, 
the valence band edge is characterized by the two energies $E_\pm$,
with halved density of states in the energy window (\ref{window2}).
One may then ask
at which energy ($E_+$ or $E_-$) the supercritical diving of a 
bound state level in an impurity potential takes place.

\section{Circular potential well}\label{sec3}

\subsection{Bound states}

\begin{figure}
\centering
\includegraphics[width=13cm]{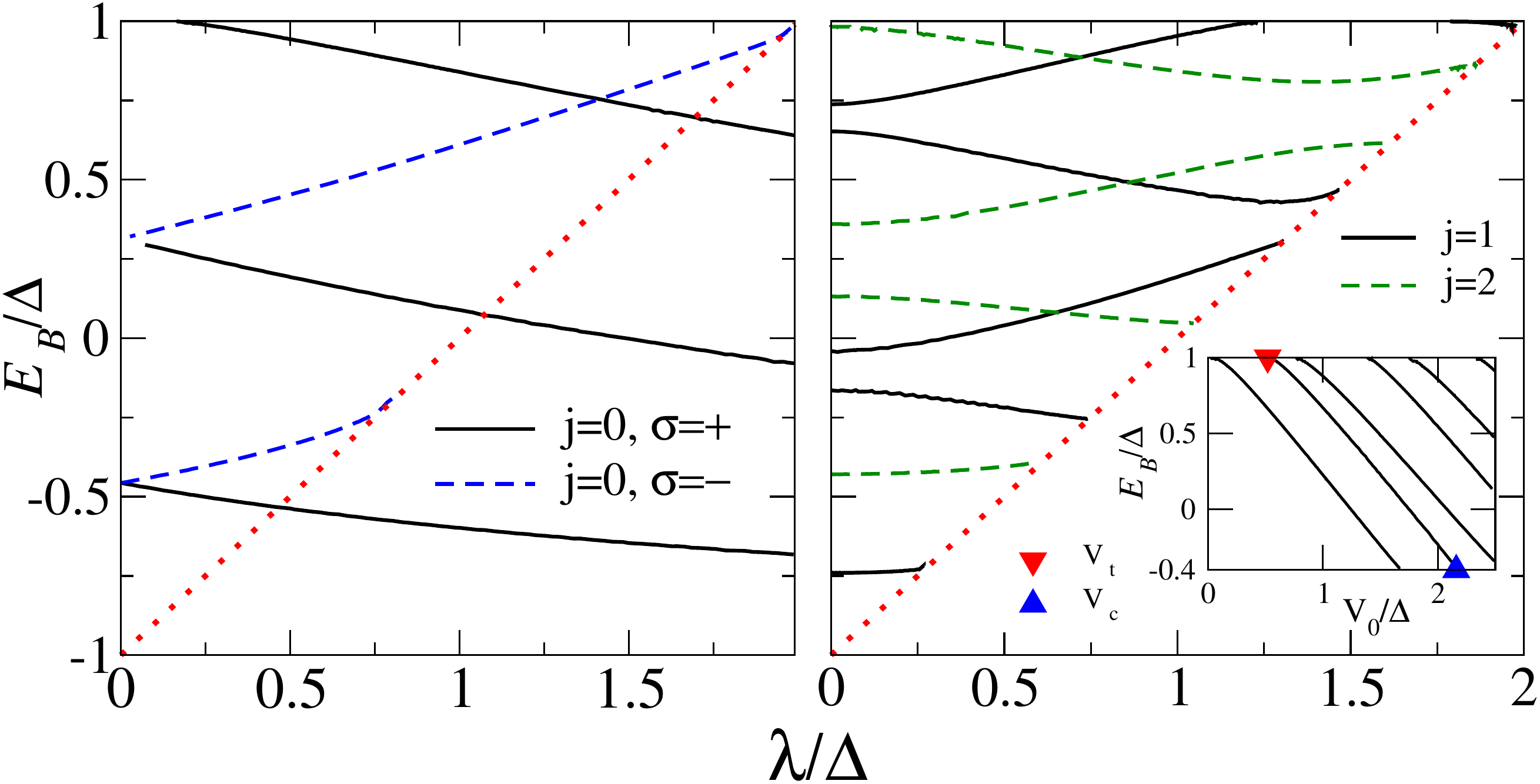}
\caption{\label{fig1} 
Bound-state spectrum ($E_B$) vs Rashba SOI ($\lambda$)  
for a circular potential well with depth $V_0=2\Delta$ and 
radius $R=3/\Delta$.  Only the lowest-energy states with $j=0,1,2$
are shown.  The red dotted line indicates $E_+=-\Delta+\lambda$.  
The left panel shows $j=0$ bound states with parity $\sigma=\pm$.
The right panel shows $j=1,2$ bound states. The
inset displays the $j=1$ bound-state energies vs potential depth $V_0$ 
for $\lambda=0.6\Delta$.  At some threshold value  $V_0=V_t$ 
(where $V_t=0$ for the lowest state shown), a new bound
state emerges from the conduction band. This state dives into
the valence band 
for some critical value $V_0=V_c>V_t$, where the valence band edge is
at energy $E_B=E_+=-0.4\Delta$. For the second bound state in the 
inset, $V_t$ ($V_c$) is shown as red (blue) triangle. 
}
\end{figure}

In this section, we study a circular potential well
with radius $R$ and depth $V_0>0$, 
\begin{equation}
V(r) = \left\{ \begin{array}{ll} -V_0, & r<R \\ 0,& r\ge R \end{array} \right.
\end{equation}
We always stay within the TI phase $\Delta>\lambda/2$,
where bound states are expected for energies $E=E_B$ in the window 
${\rm max}(\Delta-V_0,E_+)<E_B<\Delta$.
For $r<R$, the corresponding radial eigenspinor [see Eq.~(\ref{4spinor})]
is written with arbitrary prefactors $A^<_\pm$ in the form
\begin{equation} \label{ans1}
 \Phi_j^< (r)= \sum_\pm A^<_\pm\Phi_{j,\tilde p_\pm}(r),
\end{equation}
with Eq.~(\ref{generalfree2}) for $\Phi_{j,\tilde p_\pm}(r)$.
Here, the $\tilde p_\pm>0$ follow from Eq.~(\ref{ppm}) by including
the potential shift, 
\begin{equation}\label{tildeppm}
\tilde p_\pm =  (E+V_0-\Delta)(E+V_0-E_\pm).
\end{equation}
For $r>R$, the general solution is again written as
\begin{equation}\label{ans2}
\Phi_j^> (r) = \sum_\pm A^>_\pm \Phi_{j,p_\pm}(r).
\end{equation}
However, now $\Phi_{j,p_\pm}$ is given by Eq.~(\ref{Keigenstate}),
since $p_\pm<0$ for true bound states with only 
evanescent states outside the potential well.  

The continuity condition for the four-spinor at the potential step, 
$\Phi_{j}^<(R)= \Phi_{j}^>(R),$
then yields a homogeneous linear system of equations for the four 
parameters ($A^{<,>}_\pm$).  A nontrivial solution is only possible
when the determinant of the corresponding $4\times 4$ matrix $C(E)$ 
(which is too lengthy to be given here but follows directly
from the above expressions) vanishes,
\begin{equation} \label{contcond}
{\rm det} [C(E)] = 0.
\end{equation}
Solving the energy quantization condition (\ref{contcond}) 
then yields the discrete bound-state spectrum ($E_B$).
It is then straightforward to determine 
the corresponding spinor wavefunctions.

Numerical solution of Eq.~(\ref{contcond}) 
yields the bound-state spectrum  shown in Fig.~\ref{fig1}.  
When $V_0$ exceeds a ($j$-dependent) ``threshold'' value, $V_t$, 
a bound state splits off the conduction band edge. 
When increasing $V_0$ further,
this bound-state energy level moves down almost linearly, 
cf.~inset of Fig.~\ref{fig1},
and finally reaches the valence band edge $E_+=-\Delta+\lambda$
at some ``critical'' value $V_0=V_{c}$.  (For $j=0$, we will 
see below that this definition needs some revision.)
Increasing $V_0$ even further, the bound state is  then expected to 
dive into the valence band and
become a finite-width supercritical resonance, i.e.,
the energy would then acquire an imaginary part. 

\subsection{Zero angular momentum states}

Surprisingly, for  $j=0$, we find a different scenario
where supercritical diving, with finite lifetime of the resonance,
happens only for half of the bound states entering
 the energy window (\ref{window2}). Noting that
states with different parity $\sigma=\pm$ do not mix,
see Sec.~\ref{sec22}, we observe that
all $\sigma=+$ bound states enter the valence band as true
bound states (no imaginary part) throughout the energy window (\ref{window2})
while the valence band continuum is spanned by the $\sigma=-$ states.
We then define $V_c$ for $(j=0,\sigma=+)$ bound states as 
the true supercritical threshold where $E_B=E_-=-\Delta-\lambda$.
However, the $(j=0,\sigma=-)$ bound states become supercritical already 
when reaching $E_+=-\Delta+\lambda$.  

Therefore an intriguing physical situation arises for $j=0$ in the 
energy window (\ref{window2}). While $\sigma=+$ states are true
bound states (no lifetime broadening), they coexist with 
$\sigma=-$ states which span the valence band continuum or 
possibly form supercritical resonances.  For $E<E_-$, however,
all bound states dive, become finite-width resonances, and 
eventually become dissolved in the continuum.

\subsection{Threshold for bound states}

Returning to arbitrary total angular momentum $j$,
we observe that whenever $V_0$ hits a possible
threshold value $V_t$, a new bound state
is generated, which then dives into the valence band at another potential
depth $V_0=V_{c}$ (and so on). Analytical results for
all possible threshold values $V_t$ follow by expanding 
Eq.~(\ref{contcond}) for weak dimensionless binding energy 
$\delta\equiv 1-E_B/\Delta.$  For $\delta\ll 1$ and $j=1$,
Eq.~(\ref{contcond}) yields after some algebra
\begin{eqnarray}\label{disc1}
 \delta &=& \frac{2(\hbar v_F/R)^2 e^{-2\gamma} }
{\Delta^2 \sqrt{1-\tilde\lambda^2}}
\left(\frac{1-\tilde\lambda}{1+\tilde \lambda}\right)^{\tilde \lambda/2} 
e^{-\frac{(\hbar v_F/R)^2}{2V_0\Delta}
\sum_\pm z_\pm J_0(z_\pm)/J_1(z_\pm)} ,  \\ \nonumber
z_\pm &=& \sqrt{V_0(2\Delta \pm \lambda+V_0)} R ,
\end{eqnarray}
where $\gamma\approx 0.577$ is the Euler constant and  
$\tilde \lambda=\lambda/2\Delta$.
The binding energy approaches zero for $V_0\to 0$, where Eq.~(\ref{disc1})
simplifies to 
\begin{equation}\label{disc}
\delta= \frac{2(\hbar v_F/R)^2 }{\Delta^2 \sqrt{1-\tilde\lambda^2}}
\left(\frac{1-\tilde\lambda}{1+\tilde \lambda}\right)^{\tilde \lambda/2} 
e^{-2\gamma - 2 \frac{(\hbar v_F/R)^2}{V_0\Delta}}.
\end{equation}
For vanishing Rashba SOI $\lambda=0$, this reproduces 
known results \cite{gamayun2}. 
For any $\lambda<2\Delta$, we observe
that the $j=1$ bound state in Eq.~(\ref{disc}) exists for arbitrarily 
shallow potential depth $V_0$.  

The threshold values $V_t$ for higher-lying $j=1$ bound states   
also follow from the binding energy (\ref{disc1}), since
 $\delta$ vanishes for $J_1(z_+)=0$ and for $J_1(z_-)=0$. 
When one of these two conditions is
fulfilled at some $V_0=V_t$, a new bound state appears for potential depth
above $V_t$.  This statement is in fact quite general:
By similar reasoning, we find that the threshold values $V_t$ for $j=0$
follow by counting the zeroes of $J_0(z_\pm)$.
Without SOI, this has also been discussed in \cite{piet}.
Note that this argument immediately implies that no bound state with 
$j=0$ exists for $V_0\to 0$.  

From the above equations, we can then infer the threshold values $V_{t}$ 
for all bound states with $j=0$ or $j=1$ in analytical form. 
These are labeled by $n=1,2,\ldots$ and
$\sigma=\pm$ (for $j=0$, $\sigma$ corresponds to parity), 
\begin{equation}
V_{t,j,n,\pm} = (\Delta\pm \lambda/2)
\left [ 
-1 + \sqrt{1+ \gamma^2_{j,n}/[R(\Delta\pm \lambda/2)]^2} \right],
\end{equation}
where $\gamma_{j,n}$ is the $n$th zero of the $J_j$ Bessel function.

Likewise, for $j>1$, the
condition for the appearance of a new bound state is 
\begin{equation}\label{newstate}
\sum_\pm \left[ 2(j-1) z_\pm J_{j-1}(z_\pm) 
- \frac{(2\Delta\pm \lambda)V_0}{(\hbar v_F/R)^2} 
J_j(z_\pm) \right] J_j(z_\mp) = 0.
\end{equation}
Close examination of this condition shows that no 
bound states with $j>1$ exist for $V_0 \to 0$. We conclude that
bound states in a very weak potential well exist only for $j=1$.

\subsection{Supercritical behavior}\label{sec33}

\begin{figure}
\centering
\includegraphics[width=13cm]{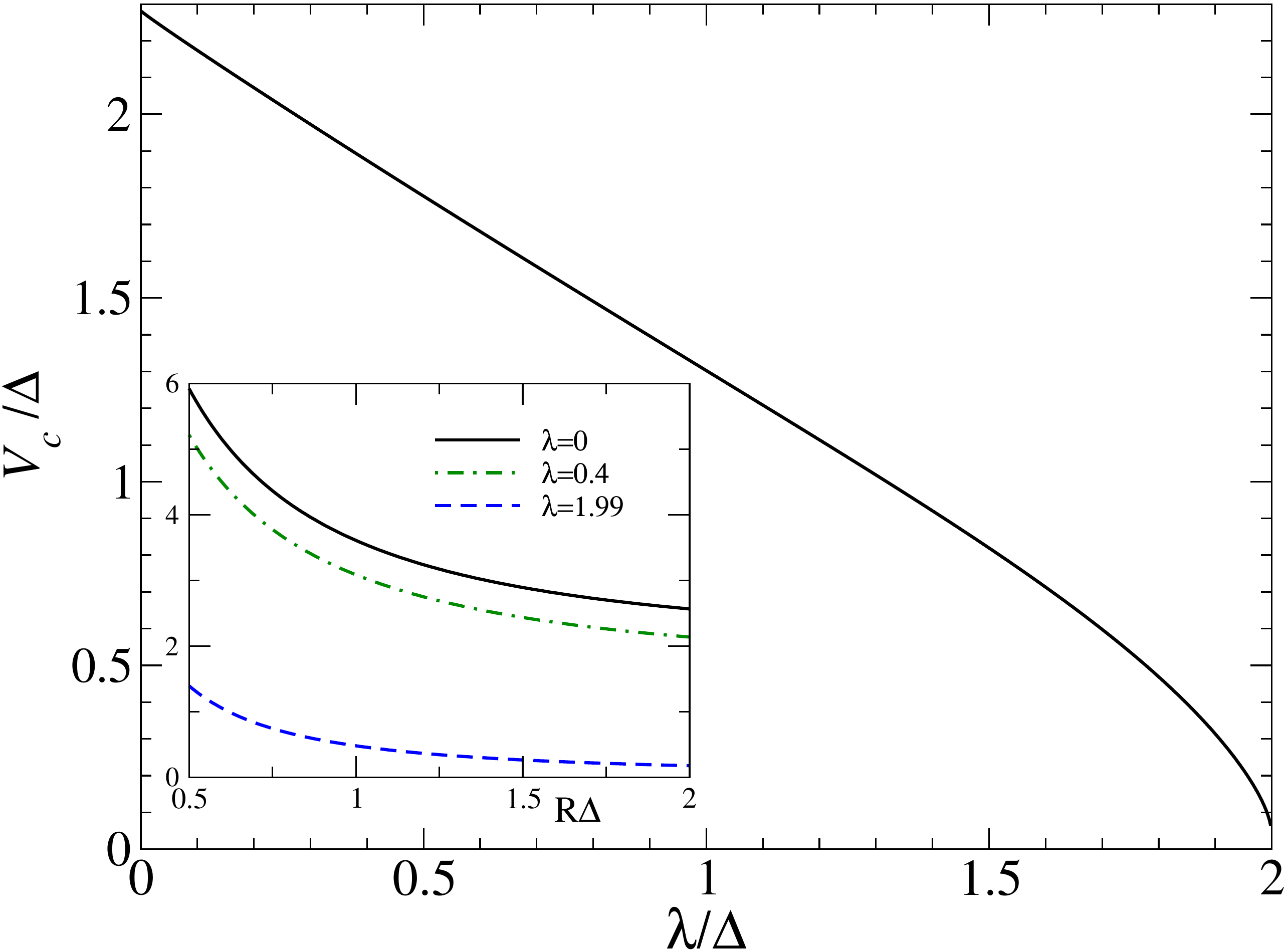}
\caption{\label{fig2}  
Critical potential depth $V_c$ for the lowest
$j=1$ bound state level in a disk with $R\Delta=3$.
The obtained $\lambda=0$ value 
matches the analytical prediction 
$V_c\approx 2.28\Delta$ from Eq.~(\ref{v0crl0}), while $V_c\to 0$ 
near the border of the TI phase ($\lambda\to 2\Delta$). 
Inset: $V_c$ vs radius $R$ with several values of $\lambda$ (given in
units of $\Delta$)
for the lowest bound state.
}
\end{figure}

As can be seen in Fig.~\ref{fig1}, the lowest $j=1$ bound state 
is also the first to enter the valence band continuum for $V_0= V_{c}$. 
For $\lambda=0$, the critical value is known to be \cite{gamayun2}
\begin{equation}\label{v0crl0}
V_{c}= \Delta \left(1 +\sqrt{1+\gamma^2_{0,1}/[R\Delta]^2}\right).
\end{equation}
with $\gamma_{0,1}\approx 2.41$.
The energy of the resonant state 
acquires an imaginary part for $V_0>V_c$ \cite{gamayun2}.
For $\lambda>0$, we have obtained implicit expressions 
for $V_{c}$, plotted in Fig.~\ref{fig2}.  Note that these
results reproduce Eq.~(\ref{v0crl0}) for $\lambda \to 0$.  
The almost linear decrease of $V_{c}$ with increasing 
$\lambda$, see  Fig.~\ref{fig2}, can be rationalized by noting that the
valence band edge is located at $E_+=-\Delta+\lambda$.  
Thereby supercritical resonances could be reached already for lower
potential depth by increasing the Rashba SOI.
Similarly, with increasing disk radius $R$, the critical value $V_c$ decreases,
see the inset of Fig.~\ref{fig2}.  
For the lowest $(j=0,\sigma=\pm)$ bound state, 
the critical value in fact follows in analytical form,
\begin{equation}
V_{c,\sigma=\pm}= (\Delta\pm \lambda/2) \left(
1+ \sqrt{1+\gamma^2_{1,1}/[(\Delta\pm\lambda/2)R]^2}\right),
\end{equation}
where $\gamma_{1,1}\approx 3.83$.

Since the parity decoupling in Sec.~\ref{sec22} only holds for $j=0$,
it is natural to expect that all $j\ne 0$ bound states turn 
into finite-width resonances when 
$E_B<E_+$.  This expectation is confirmed by an explicit calculation as follows.
Within in the window $E_-<E_B<E_+$, a true bound state should not
receive a contribution from $\Phi_{j,p_+>0}(r)$ for $r>R$, but instead has to
be obtained by matching an Ansatz as in Eq.~(\ref{ans1}) for the
spinor state inside the disk ($r<R$) to an evanescent spinor state 
$\propto \Phi_{j,p_-<0}(r>R)$.  
However, the  matching condition
is then found to have no real solution $E_B$, i.e., there
are no true bound states with $j\ne 0$ in the energy window (\ref{window2}).
We therefore conclude that all $j\ne 0$ bound states turn
supercritical when  $E_B<E_+$.  Note that
this statement includes the lowest-lying bound state (which has $j=1$).
This implies that a finite Rashba SOI can considerably lower the 
potential depth $V_c$ required for entering the supercritical regime.

\section{Coulomb center} \label{sec4}

\begin{figure}
\centering
\includegraphics[width=11cm]{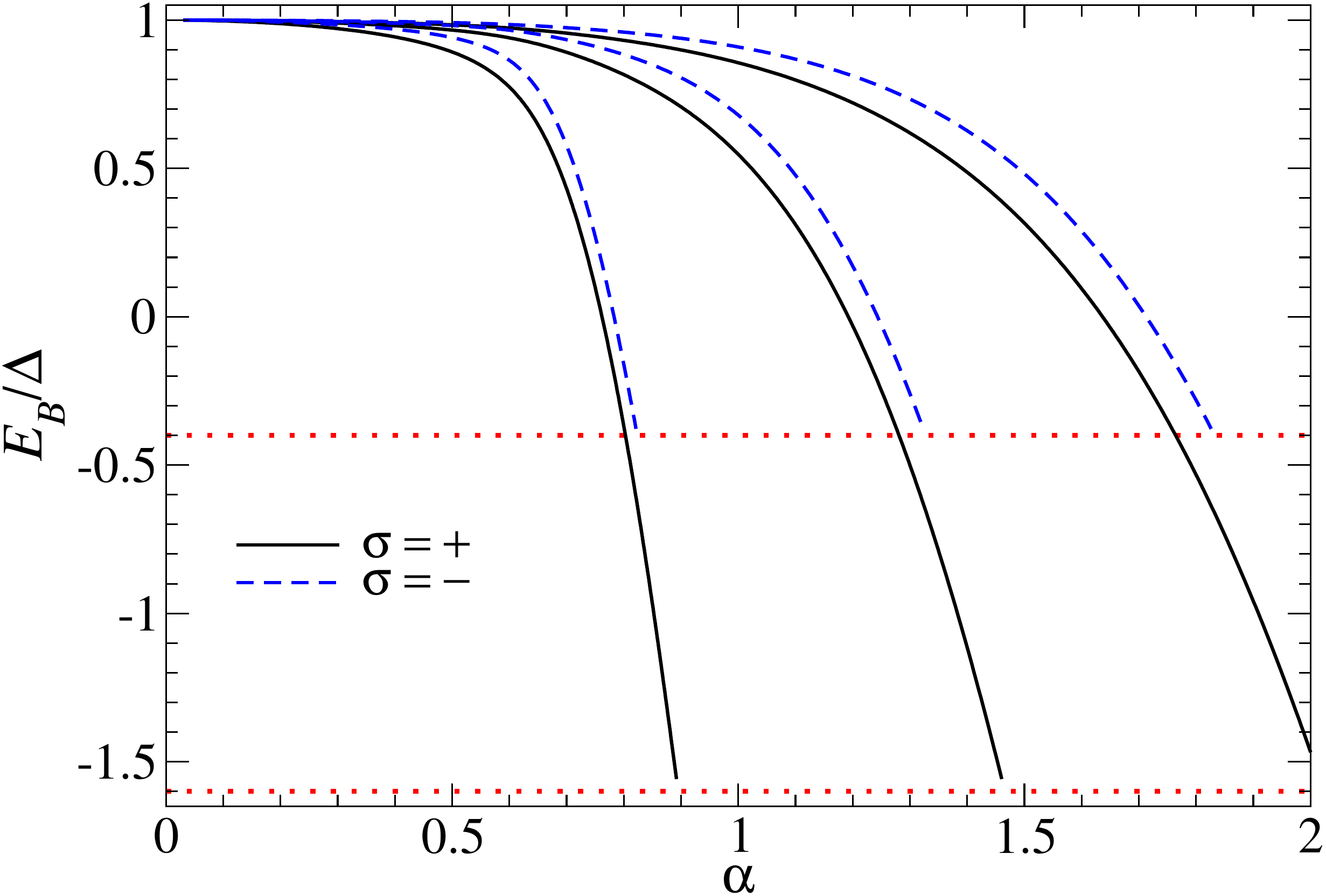}
\caption{\label{fig3} 
Bound state energies with angular momentum $j=0$ ($E_B$ in units of 
$\Delta$) vs dimensionless impurity strength $\alpha$ for the Coulomb
problem with regularization parameter $R\Delta=0.01$ and Rashba
SOI $\lambda=0.6\Delta$.  Solid black  (dashed blue)
curves correspond to parity $\sigma=+$ ($\sigma=-$). 
Results for radial number $n=1,2,3$ (with increasing energy) are shown. 
Red dotted lines denote $E=E_\pm$.
}
\end{figure}

We now turn to the  Coulomb potential, 
$V(r)=-\alpha /r$, generated by a positively charged impurity
located at the origin, with the dimensionless
coupling strength $\alpha$ in Eq.~(\ref{alphadef}).
We consider only the TI phase $\Delta>\lambda/2$ and
analyze the bound-state spectrum and conditions for
supercriticality. Again, without loss of generality,
we focus on the $K$ point only ($\tau=+$), and first summarize the known
solution for $\lambda=0$
\cite{kotov,novikov,gamayun}.  In that case, 
$s_z=\pm$ is conserved, and the spin-degenerate bound-state energies are
labeled by the integer angular momentum $j$ and a 
radial quantum number $n=1,2,3, \ldots$ (for $j>0$, $n=0$ is also possible),
\begin{equation}\label{hydr}
E_{j,n}(\lambda=0) =  \Delta\left(1+
\frac{\alpha^2}{\left(n+\sqrt{(j-1/2)^2-\alpha^2}\right)^2}\right)^{-1/2}.
\end{equation}
The corresponding eigenstates then follow in terms of hypergeometric
functions.
The lowest bound state is $E_{j=1,n=0}=\Delta\sqrt{1-4\alpha^2}$, which 
dives when $\alpha=\alpha_c=1/2$; note that
$\alpha_c$ precisely corresponds to $V_c$ in Sec.~\ref{sec3}.
In particular, for ($j=0,\sigma=\pm)$ states
 we define $\alpha_c$ in the same manner.
Next we discuss how this picture is modified when the Rashba coupling
$\lambda$ is included.

\begin{figure}
\centering
\includegraphics[width=11cm]{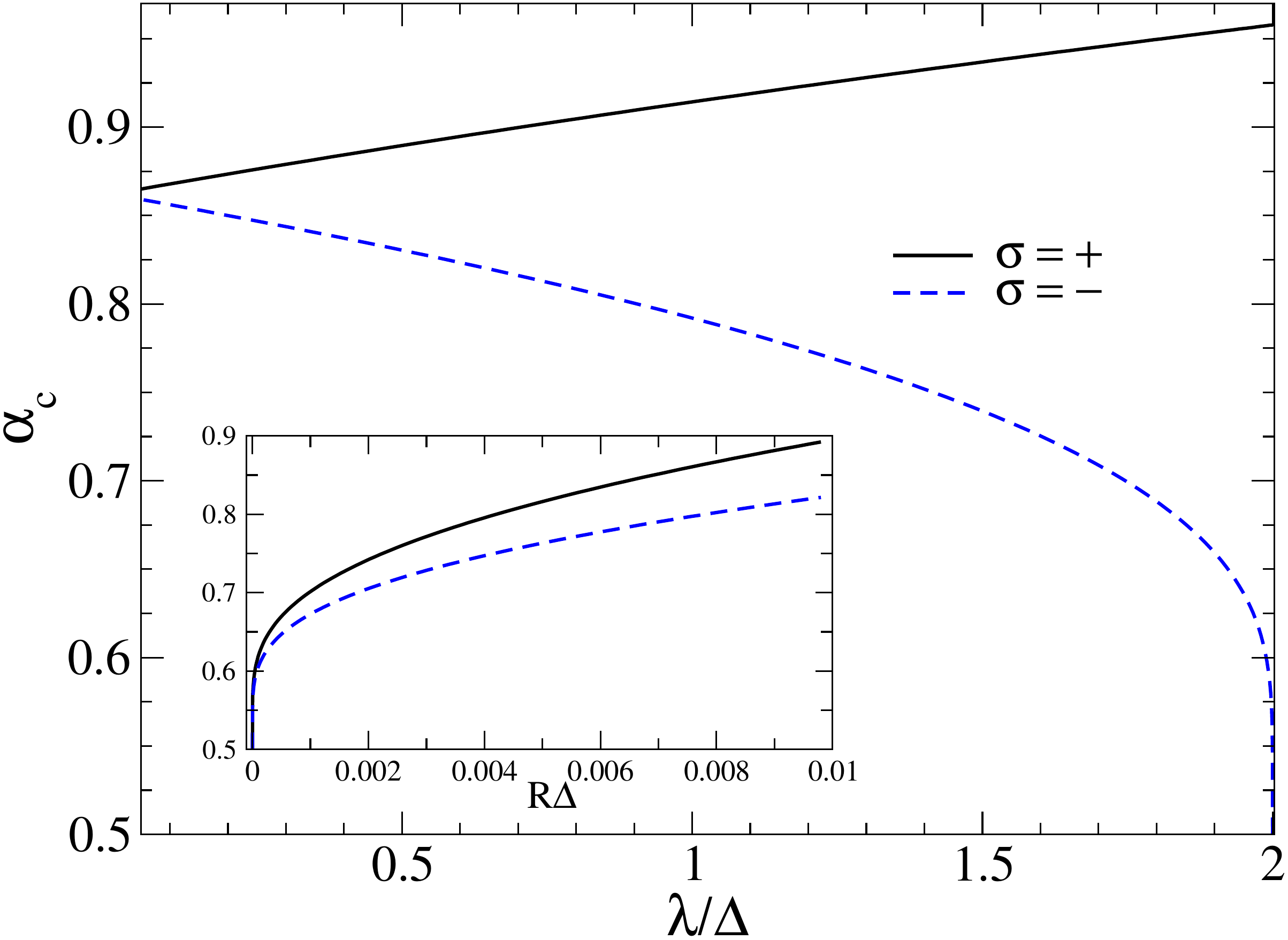}
\caption{\label{fig4} 
Main panel:
Critical Coulomb impurity strength $\alpha_c$ vs Rashba SOI $\lambda$ for 
$R\Delta=0.01$ and the lowest $(j=0,\sigma=\pm)$ bound states.
Inset: $\alpha_c$ vs cutoff scale $R$ for $\lambda=0.6 \Delta$.
}
\end{figure}

Following the arguments in Sec.~\ref{sec22} for $j=0$, the combination of
Eq.~(\ref{hydr}) with Eq.~(\ref{subst}) immediately yields the exact
 bound-state energy spectrum  ($n=1,2,3,\ldots$),
\begin{equation}
E_{j=0,n,\sigma=\pm} = (\Delta\pm\lambda/2) \left(
1+\frac{\alpha^2}{\left( n+\sqrt{1/4-\alpha^2} \right)^2}\right)^{-1/2}
\mp\lambda/2.
\end{equation}
The corresponding eigenstates then also follow from 
Refs.~\cite{novikov,gamayun}.
The very same reasoning also applies to a regularized $1/r$ potential
\cite{pereira2,gamayun}, where $V(r<R)$ 
is replaced by the constant value $V=-\alpha/R$. Here, $R$ is 
a short-distance cutoff scale of the order of the lattice spacing.
The solution of the bound-state problem then requires a wavefunction
matching procedure, which has been carried out in Ref.~\cite{gamayun}.
Thereby we can already infer all bound states for $j=0$.

Figure \ref{fig3} shows the resulting $j=0$ bound-state spectrum 
vs $\alpha$ for the regularized Coulomb potential.  
Within the energy window (\ref{window2}), we again find that
states with parity $\sigma=+$ remain true
bound states that dive only for $E_B <E_-$, while $\sigma=-$ states 
show supercritical diving already for $E_B <E_+$. 
Figure \ref{fig4} shows the 
corresponding critical couplings $\alpha_{c}$ for $\sigma=\pm$, 
where the lowest $j=0$ bound state with parity $\sigma$
turns supercritical.
Note that for finite $R$ and $\lambda\to 0$, a unique 
value for $\alpha_c$ is found, while for $\lambda\ne 0$ two 
different critical values for $\alpha_c$ are found.  However,
this conclusion holds only for finite regularization parameter $R$,
i.e., it is non-universal.  As seen in the inset of Fig.~\ref{fig4}, 
in the limit $R\to 0$, both critical values for $\alpha_c$
 approach $\alpha_c=1/2$
again, which is the value found without SOI.

Finally, for $j\ne 0$, we can then draw the same qualitative conclusions as 
in Sec.~\ref{sec33} for the potential well.  In particular, we expect
that all $j\ne 0$ bound states turn supercritical when their energy
$E_B$ reaches the continuum threshold at $E_B=E_+=-\Delta+\lambda$.

\section{Conclusions}\label{sec5}

In this work, we have analyzed the bound-state problem for the 
Kane-Mele model of graphene with intrinsic ($\Delta$) and Rashba
($\lambda$) spin-orbit
couplings when a radially symmetric  attractive potential $V(r)$
is present.  We have focussed on the most interesting ``topological
insulator'' phase with $\Delta>\lambda/2$. 
The Rashba term $\lambda$ leads to a restructuring of the valence band, with
a halving of the density of states in the window $E_-<E<E_+$,
 where $E_\pm=-\Delta \pm \lambda$.
This has spectacular consequences for total angular momentum $j=0$,
where the problem can be decomposed into
two independent parity sectors ($\sigma=\pm$). The $\sigma=+$ states
remain true bound states even inside the above window and coexist
with the continuum solutions as well as 
possible supercritical resonances in the $\sigma=-$ sector.
However, all $j\ne 0$ bound states exhibit supercritical diving for $E<E_+$,
where the critical threshold ($V_c$ or $\alpha_c$ for the disk or the
Coulomb problem, respectively) is lowered when the Rashba term is present.
We hope that these results will soon be put to an experimental test.

\section*{Acknowledgements}

This work has been supported by the DFG within the 
network programs SPP 1459 and SFB-TR 12.

\bibliographystyle{mdpi}
\makeatletter
\renewcommand\@biblabel[1]{#1. }
\makeatother

\end{document}